\documentclass[]{aa}

\def\rmit#1{{\it #1}}              

\def\eg{\rmit{e.g.}}

\usepackage{booktabs}
\usepackage{graphicx}
\usepackage{multirow}
\usepackage{color}
\usepackage[flushleft]{threeparttable}

\usepackage{array}
\newcolumntype{?}{@{\vrule width 2pt}}

\usepackage{hyperref}
\hypersetup{
    final=true,
    pageanchor=true,
    colorlinks=true,
    breaklinks=true,
    linkcolor=blue,
    citecolor=blue,
    urlcolor=blue,
    pdfpagemode=UseNone,
    pdftitle={},
    pdfauthor={T. Felipe et al.},
    pdfsubject={Solar Physics},
    pdfkeywords={Methods: observational -- Sun: photosphere -- Sun: chromosphere -- Sun: oscillations  -- sunspots -- Techniques: polarimetric}}

\titlerunning{Detection of farside active regions with deep learning}

\begin{document}

\title{Improved detection of farside solar active regions using deep learning}

\author{T. Felipe\inst{\ref{inst1},\ref{inst2}}
\and A. Asensio Ramos\inst{\ref{inst1},\ref{inst2}}
}

\institute{Instituto de Astrof\'{\i}sica de Canarias, 38205, C/ V\'{\i}a L{\'a}ctea, s/n, La Laguna, Tenerife, Spain\label{inst1}
\and 
Departamento de Astrof\'{\i}sica, Universidad de La Laguna, 38205, La Laguna, Tenerife, Spain\label{inst2} 
}

\abstract
{The analysis of waves in the visible side of the Sun allows the detection of active regions in the farside through local helioseismology techniques. The knowledge of the magnetism in the whole Sun, including the non-visible hemisphere, is fundamental for several space weather forecasting applications.}
{Seismic identification of farside active regions is challenged by the reduced signal-to-noise, and only large and strong active regions can be reliable detected. Here we develop a new methodology to improve the identification of active region signatures in farside seismic maps.}
{We have constructed a deep neural network that associates the farside seismic maps obtained from helioseismic holography with the probability of presence of active regions in the farside. The network has been trained with pairs of helioseismic phase shift maps and Helioseismic and Magnetic Imager magnetograms acquired half a solar rotation later, which were used as a proxy for the presence of active regions in the farside. The method has been validated using a set of artificial data, and it has also been applied to actual solar observations during the period of minimum activity of the solar cycle 24.}
{Our approach shows a higher sensitivity to the presence of farside active regions than standard methods applied up to date. The neural network can significantly increase the number of detected farside active regions, and will potentially improve the application of farside seismology to space weather forecasting.}
{}


\maketitle


\section{Introduction}

Helioseismology studies the solar interior by analyzing the oscillations
observed at the surface. Its first applications were based on the interpretation
of accurate measurements of the eigenfrequencies of the resonant oscillatory
modes. This field has been labeled as ``global helioseismology'', and it has
revealed the internal structure and rotation of the Sun
\citep[\eg,][]{Christensen-Dalsgaard2002}. Since the late 80s, a complementary
set of techniques and theoretical methodologies, known as ``local
helioseismology'', have been developed in order to probe local regions of the
solar interior or surface. Local helioseismology does not focus only on the
resonant modes, but studies the full wave field instead. This approach allows to
measure longitudinal variations and meridional flows, in contrast to global
helioseismology. See \citet{Gizon+Birch2005} for a review on local
helioseismology.

One of the most remarkable applications of local helioseismology is the
detection of active regions at the non-visible hemisphere of the Sun (farside).
This was first achieved using the technique of helioseismic holography
\citep{Lindsey+Braun2000, Braun+Lindsey2001}. Helioseismic holography was
introduced by \citet{Lindsey+Braun1990}. A detailed description of the method
can be found in \citet{Braun+Birch2008}. It uses the wavefield measured in a
region of the solar surface (called ``pupil'') to determine the wavefield at a
``focus point'' located at the surface or at a certain depth. This inference is
performed assuming that the observed wavefield at the pupil (\eg, the
line-of-sight Doppler velocity) is produced by waves converging toward the focus
point or waves diverging from that point. Farside helioseismic holography is a
particular application of this methodology, where the pupil is located at the
nearside hemisphere and the focus points are located at the surface in the
farside hemisphere \citep[see][for a thorough discussion of this
technique]{Lindsey+Braun2017}. The identification of active regions is founded
on the fact that they introduce a phase shift between ingoing and outgoing waves
\citep{Braun+etal1992}. This phase shift (which can be characterized as a
travel-time shift) is mainly due to the depression of the photosphere in
magnetized regions, which produces that the upcoming waves reach the upper
turning point a few seconds earlier in active regions than in quiet Sun regions
\citep{Lindsey+etal2010, Felipe+etal2017b}. This way, when an active region is
located at the focus point, a negative phase shift (reduction in the travel
time) is found. Farside imaging has later been performed using time-distance
helioseismology \citep{Duvall+Kosovichev2001,Zhao2007, Ilonidis+etal2009}.  

\begin{figure*}[!ht] 
 \centering
 \includegraphics[width=18cm]{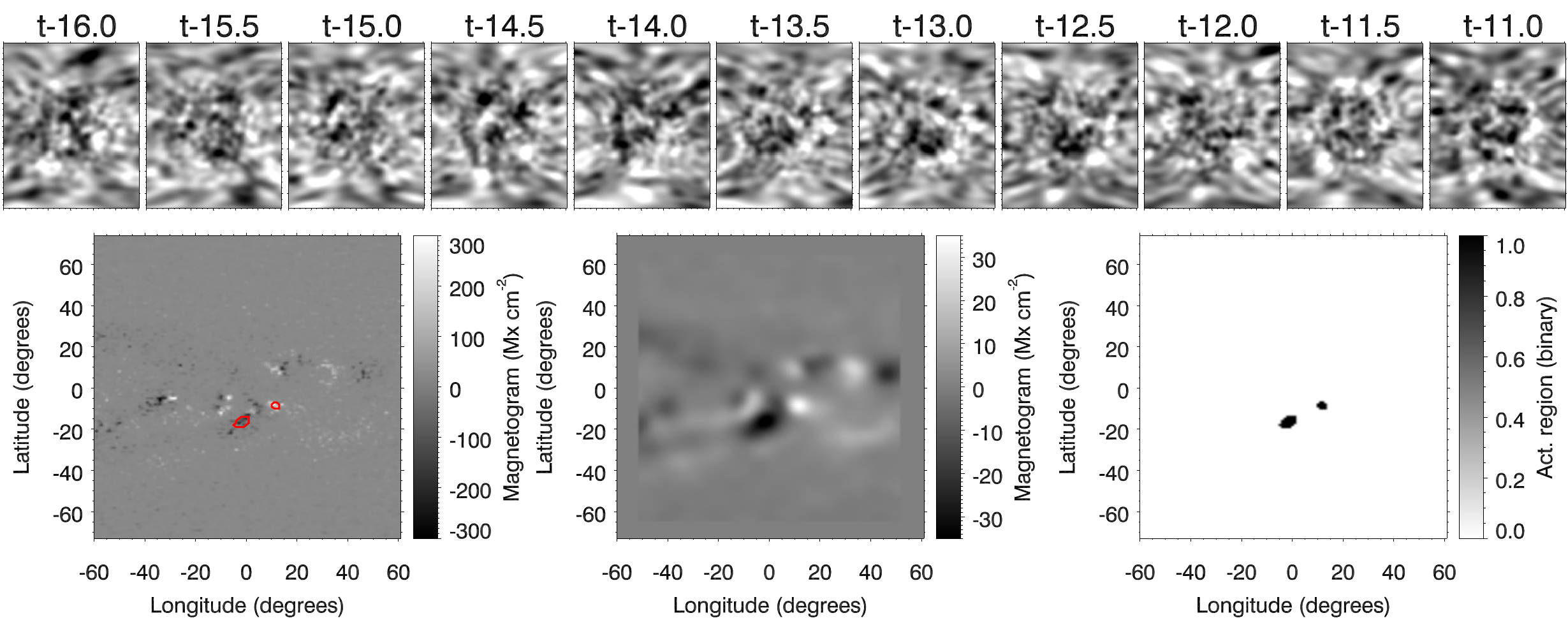}
  \caption{Example of one of the elements from the training set. Panels in the top row show 11 farside seismic maps, each of them obtained from the analysis of 24 hours of HMI Doppler data. The horizontal axis is the longitude (a total of 120$^{\circ}$) and the vertical axis is the latitude (between -72$^{\circ}$ and 72$^{\circ}$). The label above the panels indicates the number of days prior to the time $t$ when the corresponding magnetogram was acquired (in this example, $t$ is 2015 December 10 at 12:00 UT). The bottom row shows the magnetograms used as a proxy for the presence of active regions: Left panel: original magnetogram in heliospheric coordinates; middle panel: magnetogram after removing active regions emerged in the nearside and applying a Gaussian smoothing; right panel: binary map, where a value of $1$ indicates the presence of an active region in those locations whose magnetic flux in the smoothed magnetogram is above the selected threshold. Red contours in bottom left panel delimit the regions where the binary map is $1$. The neural network is trained by associating the 11 farside seismic maps (top row) with the binary map.}      
  \label{fig:training_example}
\end{figure*}

Farside maps computed using helioseismic holography are routinely calculated
twice a day using Doppler velocity wavefields obtained in 24 hours windows. They
are archived and accessible through the internet. Those maps are measured from
GONG data\footnote{\url{https://farside.nso.edu}} and Helioseismic and Magnetic
Imager (HMI) data\footnote{\url{http://jsoc.stanford.edu/data/farside}}. The
interest in the detection of active regions in the farside goes beyond the
simple curiosity of measuring them before they rotate into the visible
hemisphere. The knowledge of the magnetism in the whole Sun (including the
non-visible hemisphere) is fundamental for several space weather forecasting
applications. One of them is the forecasting of the UV and EUV irradiance on
Earth, since active regions have a strong impact on the irradiance at those
wavelengths. \citet{Fontenla+etal2009} showed that including the information of
the helioseismic farside maps significantly improves the Ly$\alpha$ irradiance
forecasting. This method can be extended to forecast the entire FUV and EUV
irradiance spectrum. Data driven photospheric flux transport models including
active regions in the farside also improve the solar wind forecast and the F10.7
index (solar radio flux at 10.7 cm) forecast \citep{Arge+etal2013} and allow to
successfully estimate the location and magnitude of large active regions before
they are visible in the nearside \citep{Schrijver+DeRosa2003}. Models including
farside detection of active regions have also been used to explore the open flux
problem, that is, the discrepancy between the magnetic flux in open field
regions of the Sun and that measured in situ by spacecrafts
\citep{Linker+etal2017}. 

One of the main limitations of farside helioseismology is the reduced
signal-to-noise. The signature of an active region detected in the farside has a
signal-to-noise around 10, which makes that only large and strong active regions
can be reliable detected in farside phase shift maps \citep[several hundred of
active regions per solar cycle,][]{Lindsey+Braun2017}. The goal of this paper is
to improve the identification of active region signatures in farside phase-shift
maps using a deep learning approach. The paper is organized as follows: Sect.
\ref{sect:network} describes the neural network, including the data employed for the training set, Sect. \ref{sect:tests} shows the evaluation of the
performance of our methodology using artificial data sets, Sect.
\ref{sect:observations} presents the results from the application of the network
to actual solar data, and, finally, in Sect. \ref{sect:conclusions} we discuss
the results and draw the conclusions.      

\begin{figure*}[!ht] 
  \centering
  \includegraphics[width=\textwidth]{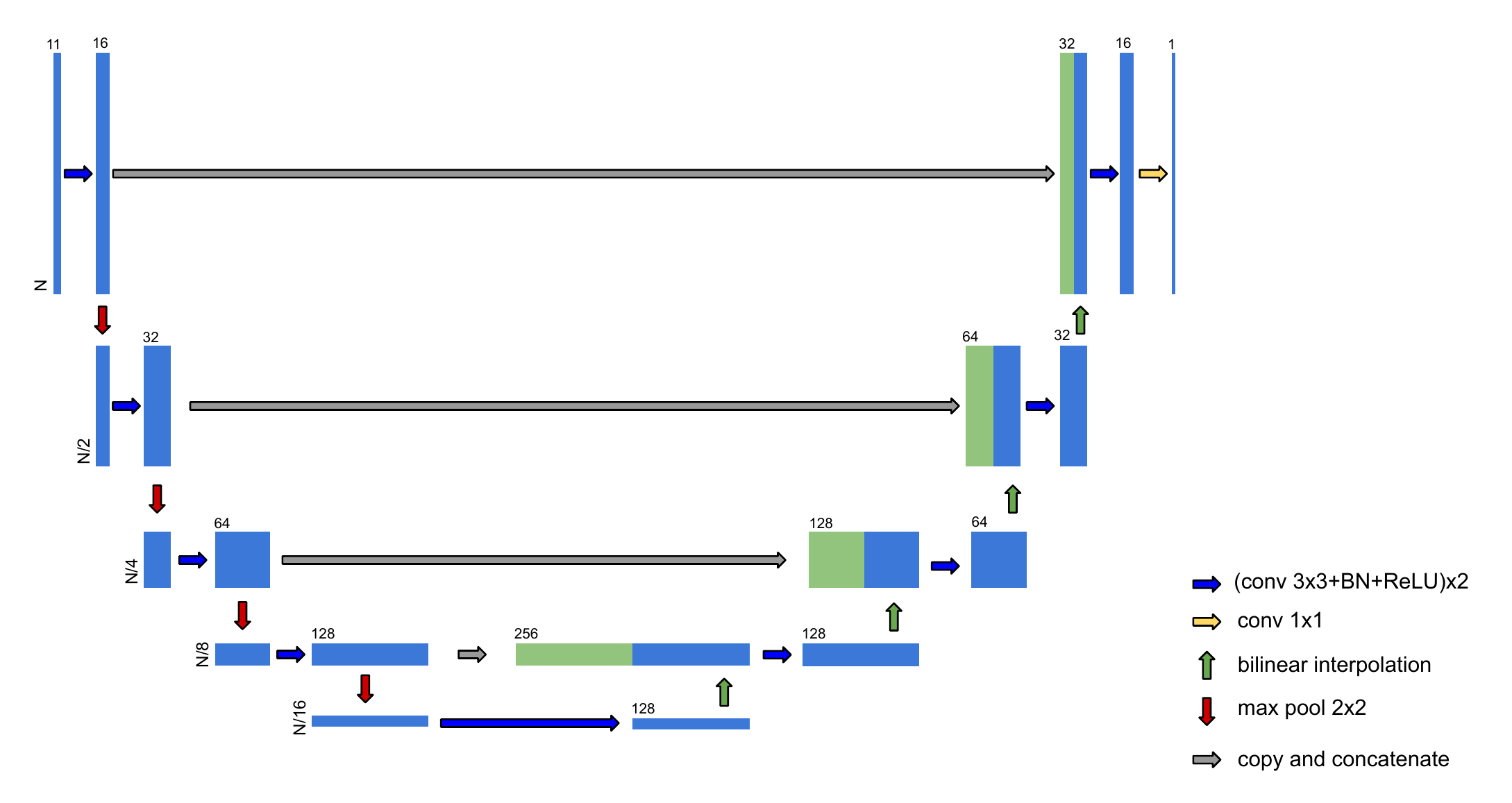}
   \caption{U-net architecture. The vertical extent of the blocks indicate the
   size of the image, while the numbers above each block shows the number of
   channels.}
   \label{fig:U-net}
 \end{figure*}

\section{Neural network approach}
\label{sect:network}
Indubitably, the recent success of machine learning is a consequence of
our ability to train very deep neural networks 
\citep[DNNs; see][]{Goodfellow-et-al-2016}. 
DNNs can be seen as a very flexible and differentiable parametric mapping 
between an input space and an output space. These highly parameterized
DNNs are then tuned by optimizing a loss function, which measures
the ability of the DNN to map the input space onto the output space
over a predefined
training set. The combination of loss function and specific
architecture has to be chosen to solve the specific problem at hand.

Arguably, the 
largest number of applications of DNNs has been in computer 
vision\footnote{See, e.g., the curation on \texttt{\url{https://bit.ly/2ll0dQI}}.}.
Problems belonging to the realm of machine vision can hardly
be solved using classical methods, be it based on machine learning or rule-based 
methods. 
Only now, with the application of very deep neural networks, have we been 
able to produce real advances.
Applications in science, and specifically in astrophysics and solar physics,
have leveraged the results of machine vision to solve problems that were 
difficult or impossible to deal with in the past with classical techniques.
The literature is growing very fast but, as a summary, we find applications 
ranging from the classification of galactic 
morphologies \citep{huertas-company15} or the development of generative 
models to help constrain the deconvolution of images of 
galaxies \citep{schawinski17} to the real-time multiframe blind deconvolution
of solar images \citep{2018A&A...620A..73A} or the probabilistic
inversion of flare spectra \citep{2019ApJ...873..128O}. Our aim in this
work is to apply convolutional neural networks to learn a very fast
and robust mapping between consecutive maps of estimated seismic maps 
and the probability map of the presence of an active region on the farside.

\subsection{Training set}
\label{sect:data}

We have designed a neural network that can identify the presence of active
regions in the farside. As input, the network uses farside phase-shift maps
computed using helioseismic holography. As a proxy for the presence of active
regions, we employed HMI magnetograms measured in the nearside (facing Earth).
The details of the data are discussed in the following sections. 
The training set that we describe in this section is used to supervisedly tune the
parameters of the neural network with the aim of being generalizable to 
new data.

\begin{figure*}[!ht] 
 \centering
 \includegraphics[width=16cm]{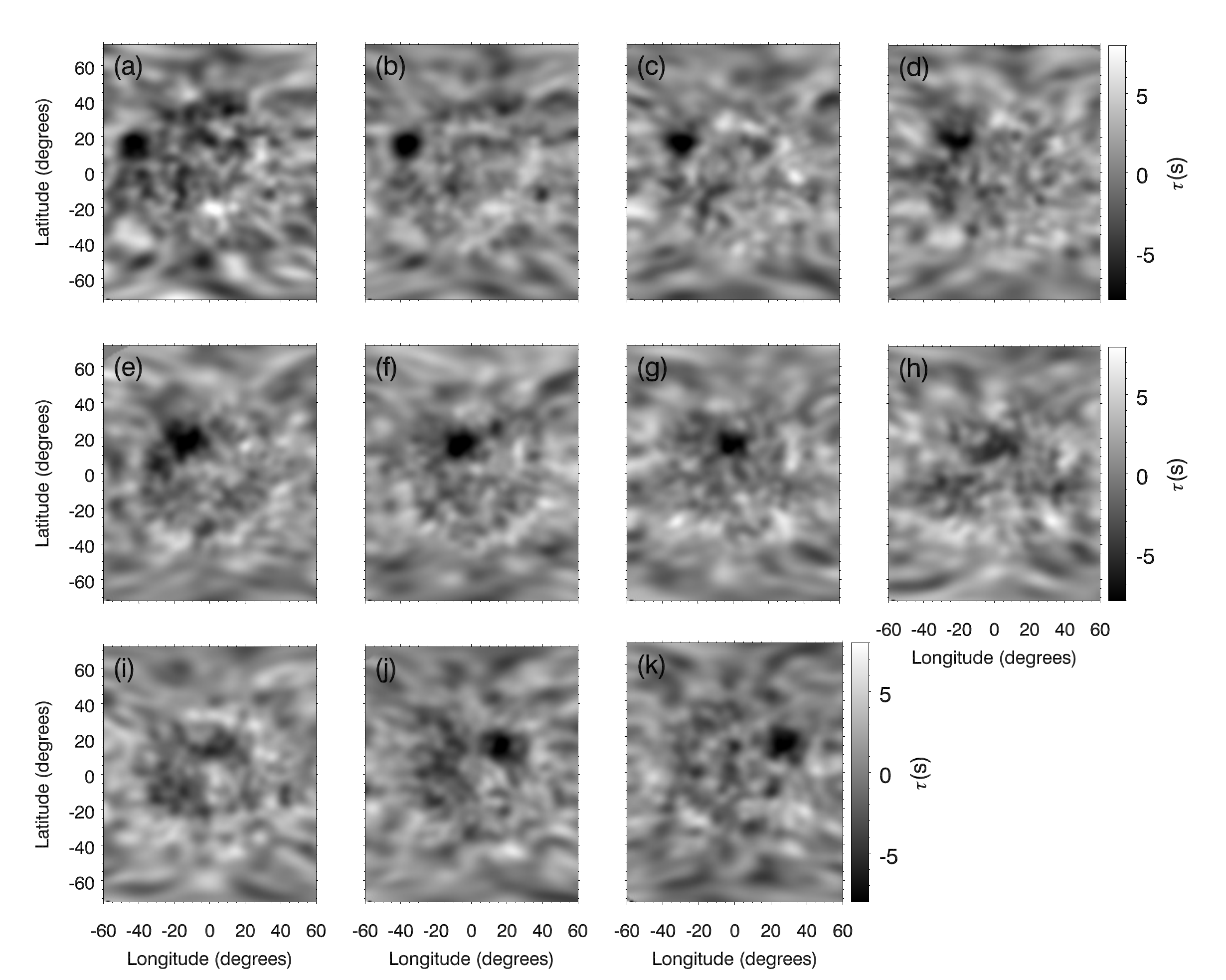}
  \caption{Artificial seismic maps for an acoustic source with $A=-9$ s, $\mathrm{FWHM}=15^{\circ}$, and a latitude of $15^{\circ}$. Time increases from panel a to panel k, with a temporal step of 12 hours.  }      
  \label{fig:11maps}
\end{figure*}

\subsubsection{HMI magnetograms}

HMI Magnetograms are one of the data products from the Solar Dynamics Observatory
available through the Joint Science Operations Center (JSOC). In order to
facilitate the comparison with the farside seismic maps (next section), we are
interested in magnetograms remapped onto a Carrington coordinate grid. We used data from the JSOC series \texttt{hmi.Mldailysynframe\_720s}. This data series contains synoptic maps constructed of HMI magnetograms collected over a 27-day solar rotation, where the first 120 degrees in longitude are replaced by data within 60 degrees of the central meridian of the visible hemisphere observed approximately at one time. These
maps are produced daily at 12 UT. We only employed the 120 degrees in longitude
including the magnetogram visible on the disk at one time. Magnetograms between
2010 June 1 (the first date available for the \texttt{hmi.Mldailysynframe\_720s}
data) and 2018 Oct 26 were extracted. Since one magnetogram is taken per day,
they make a total of 3066 magnetograms. Figure \ref{fig:training_example} summarizes the data employed for the training set. One of the original magnetograms in heliospheric coordinates is shown at the bottom left panel. 

Due to the emergence of new active regions and the decaying of the old ones,
magnetograms obtained in the nearside are an inaccurate characterization of the
active regions present in the farside half a rotation earlier or later. We have
partially corrected this issue. The farside maps are associated to the
magnetogram obtained when the seismically probed region has fully rotated to the
Earth-side, that is, 13.5 days after the measurement of the farside map. We have
removed the active regions that emerge in the nearside, since they were not
present when the farside seismic data was taken. In order to identify the
emerging active regions, we have employed the Solar Region Summary (SRS)
files\footnote{Available at \url{ftp.swpc.noaa.gov/pub/warehouse/}}, where the
NOAA registered active regions are listed. All the active regions that appear
for the first time at a longitude greater than -60 degrees (where the 0
corresponds to the central meridian of the visible hemisphere and the minus sign
indicate the east hemisphere) were masked in the magnetograms. The value of the
magnetogram was set to zero in an area 15 degrees wide in longitude and 12
degrees wide in latitude, centered in the location of the active region reported
in the SRS file of that date (after correcting the longitude, since we are
employing magnetograms retrieved at 12 UT and in the SRS files the location of
the active regions are reported for 24 UT on the previous day). Those active
regions that emerge in the visible hemisphere too close to an active region that
had appeared on the east limb due to the solar rotation were not masked. From the total of 1652 active regions labeled by NOAA during the temporal period employed for the training set, 967 were masked since they emerge in the visible hemisphere.   

The neural network is trained with binary maps,
where the 0s correspond to quiet regions and the 1s to active regions. This 
binary mask is built from the corrected magnetograms as follows. A
Gaussian smoothing with a standard deviation of 3 degrees has been applied to
the corrected magnetograms. This smoothing removes all small-scale
activity in the map and facilitates the segmentation of active regions of 
importance in the magnetogram.
Then, those regions with a magnetic flux higher than
30 Mx cm$^2$ are identified as active regions (and set to 1), and regions with lower magnetic flux are set to 0. The middle panel from the bottom row from Fig. \ref{fig:training_example} shows the magnetogram after removing the active regions emerged in the visible solar hemisphere and applying the Gaussian smoothing. The active region visible in the original magnetogram (bottom left panel from Fig. \ref{fig:training_example}) at a longitude $-30^{\circ}$ and a latitude $-5^{\circ}$ emerged in the nearside and, thus, it was masked. The bottom right panel of Fig. \ref{fig:training_example} shows the binary map indicating the location of the remaining active regions, those whose magnetic flux is above the selected threshold. Note that their positions match that of some regions with strong negative travel times in the seismic maps from about half a rotation earlier (case ``t-13.0'' in the top row from Fig. \ref{fig:training_example}).

\subsubsection{Farside phase-shift maps}
\label{sect:farside_maps}
Phase-shift maps of the farside region of the Sun are available through JSOC. They are computed from
HMI Doppler data using temporal series of one or five days. The processing of
series of 5 days is a novel approach since 2014, introduced to improve the
signal-to-noise of the phase-shift maps. They are provided in Carrington
heliographic coordinates with a cadence of 12 hours (maps are obtained at 0 and
12 UT). In this work, we have focused on the farside maps computed from 24 hours
of Doppler data. We have employed farside maps between 2010 May 18 and 2018 Oct
12. For each map, we selected a 120$^{\circ}$ region in longitude centered at
the Carrington longitude of the central meridian of the visible hemisphere 13.5
days after the date of the farside map. This way, corrected magnetograms with
the new active regions removed are associated to farside maps sampling the same
region in longitude. The training employs 11 consecutive farside maps for each corrected magnetogram, improving the seismic signal. These 11 consecutive farside maps correspond to 6 days of data. The latitude span of the maps is
between $-72^{\circ}$ and 72$^{\circ}$. We choose a sampling of 1$^{\circ}$ in both
latitude and longitude.

JSOC also reports routinely the farside active regions detected from the seismic
analysis. A detection is claimed when the phase shift integrated over an area
exceeds a certain threshold value. The area of integration is determined as a
region where the phase shift is lower than -0.085 radian. The reports with the
farside active regions are published two times a day. We have used these reports
to evaluate the performance of the neural network in comparison with the
traditional approach for detecting farside active regions (see Sect.
\ref{sect:observations}).

\subsection{Neural network architecture}
\label{sect:neural_network}

The neural network of choice in 
this work is an U-net \citep{2015arXiv150504597R}, a fully
convolutional architecture that has been used extensively for dense segmentation
of images and displayed in Fig. \ref{fig:U-net} \citep[e.g.,][in astrophysics]{2019arXiv190611248H,2019Icar..317...27S}. The U-net 
is an encoder-decoder 
network, in which the input is
successively reduced in size via contracting layers and finally increased in size via
expanding layers. This encoder-decoder architecture has three main advantages, all
of them a consequence of the contracting/expanding layers. The first
one is that the contracting layers reduce the size of the images at each step.
This makes the network faster because convolutions have to be carried out
over smaller images. The second advantage is that this contraction couples
together pixels in the input image that were far apart, so that smaller kernels
can be used in convolutional layers (we use $3 \times 3$ kernels) and the network
is able to better exploit multiscale information. The final
advantage is a consequence of the skip connections (grey 
arrows), which facilitates training by explicitly
propagating multiscale information from the contracting layers to the
expanding layers.

As shown in Fig. \ref{fig:U-net}, the specific U-net architecture
we use in this work is a combination of several
differentiable operations. The first one, indicated with blue arrows, is
the consecutive application of convolutions with 3$\times$3 kernels, 
batch normalization
\citep[BN;][]{batch_normalization15}, which normalize the input so that its mean
is close to zero and its variance close to unity (which is known to
be an optimal range of values for neural networks to work best) and
a rectified linear unit (ReLU) activation function, given 
by  $\sigma(x)=\max(0,x)$ \citep{relu10}. This combination 
Conv+BN+ReLU is repeated twice as indicated in
the legend of Fig. \ref{fig:U-net}. Red arrows refer to 
max-pooling \citep[e.g.,][]{Goodfellow-et-al-2016}, which reduces the resolution
of the images by a factor 2 by computing the maximum of all non-overlapping 
$2 \times 2$ patches in the image. The expanding layers increase again the size of the images
by using bilinear interpolation (green arrows) followed by convolutional layers.
Additionally, the layers in the encoding part transfer information to the
decoding part via skip connections (grey arrows), which greatly 
improves the ability and stability of the network.
Finally, since the output is a probability map, we force it to be in the $[0,1]$ range
thanks to a sigmoid activation function applied in the last layer after a final
$1 \times 1$ convolution used to reduce the number of channels from 16 to 1.

The neural
network is trained by minimizing the binary cross entropy between the output of
the network per pixel ($p_i$) and the binarized magnetograms ($y_i$), summed
over all pixels in the output magnetogram ($N$):
\begin{equation}
    \ell = -\frac{1}{N} \sum_{i=1}^{N} y_{i} \cdot \log p_i+
    \left(1-y_{i}\right) \cdot \log \left(1-p_i\right)
\end{equation}
For optimizing the previous loss function we employ the Adam optimizer \citep{adam14} with a
constant learning rate of 3$\times$10$^{-4}$ during 300 epochs and a batch
size of 30.

The neural network makes use of the open-source packages \texttt{numpy} \citep{numpy}, \texttt{matplotlib} \citep{matplotlib}, \texttt{astropy} \citep{astropy}, \texttt{h5py} \citep{hdf5}, \texttt{scipy} \citep{scipy}, and \texttt{PyTorch} \citep{pytorch}.

\begin{figure}[!ht] 
 \centering
 \includegraphics[width=9cm]{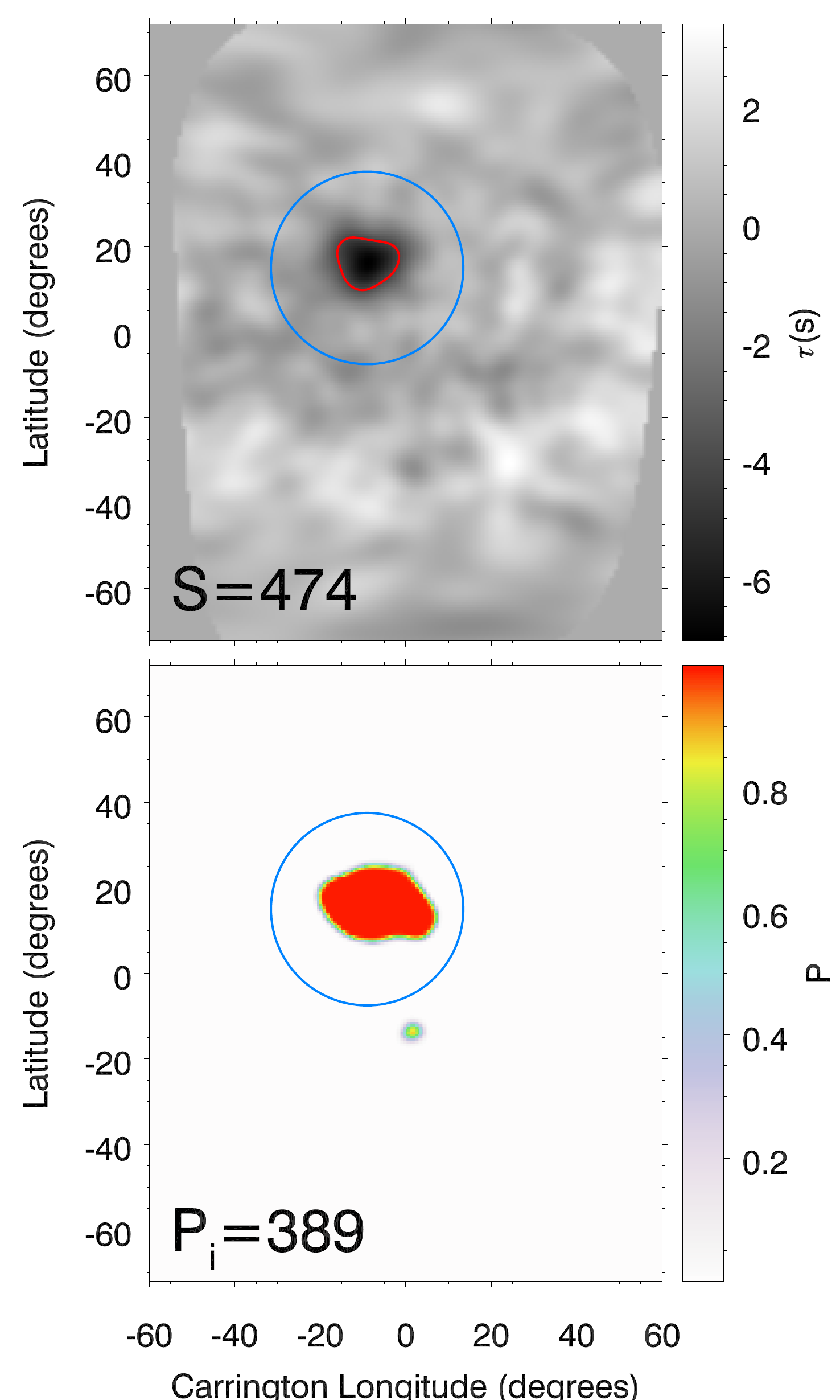}
  \caption{Top panel: 5 day average of the phase shift for the artificial case illustrated in Fig. \ref{fig:11maps}. Red contour delimits the region where the phase shift exceeds 0.085 rads. The strength of the acoustic source is shown at the bottom-left corner. Bottom panel: Probability map of the artificial active region illustrated in Fig. \ref{fig:11maps}, as retrieved from the application of the neural network. The integrated probability $P_{\rm i}$ of the feature inside the blue circle is shown at the bottom-left corner. In both panels the blue circle is centered at the location of the acoustic source, with a diameter of 3 times its FWHM. }      
  \label{fig:prediction_ejemplo}
\end{figure}

\section{Artificial tests}
\label{sect:tests}

We have evaluated the performance of the neural network using artificial maps of
farside phase shifts. These artificial maps are constructed by adding a source
(with a Gaussian shape) in the phase shift to farside seismic maps that only
contain noise. The procedure for building the artificial farside maps is the
following. First, we have selected a set of observational farside seismic maps
which do not contain any signal from active regions. They must satisfy the
following conditions: (1) They were measured around the solar minimum, in order
to minimize the chances of appearance of an active region. Maps between November
2017 and February 2019 were chosen. (2) No active region must be present in the
visible eastern (western) limb in the 14 days after (prior to) the measurement
of the seismic map. (3) The maximum magnitude of phase shift must not exceed $-8$
s. A total of 111 noise maps that satisfy these conditions were selected. 

Second, a temporal series is constructed by randomly selecting 11 maps from the
whole set of noise maps. Since the original noise maps are located at different
Carrington longitudes, they have been displaced in longitude so they resemble a
continuous series with a map measured every 12 hours. That is, the first noise
map is placed at a certain longitude, and the successive maps are centered at a
different longitude taking into account the solar rotation after half a day. The
public farside maps employed are published every 12 hours, and each of them is
measured over a 24 hours window. This way, there is a 12 hours overlap between
two consecutive maps. In order to mimic this, each randomly selected noise map
is averaged with that selected for the previous time step. We have chosen to
perform this method instead of just selecting 11 consecutive noise maps to avoid
the signal from unnoticed active regions. With this approach, if an active
region is present in the set of 111 noise maps it will have a minor impact in
the resultant noise series of 11 maps.

Third, a Gaussian phase shift perturbation (representing the perturbation of an
active region) is added to the noise maps of the temporal series. The Gaussian
perturbation is characterized by its position in longitude and latitude, its
amplitude ($A$), and its full width half maximum (FWHM). We also explored the
temporal variations of the farside signals. On the one hand, we evaluated the
lifetime of the active region through the inclusion of the Gaussian perturbation
in a certain number of consecutive days from the 11 maps that compose each case.
On the other hand, since farside active regions that are detected helioseismically 
usually show obvious day-to-day variations (even some large active regions
are not consistently visible in each image, often disappearing for one day and
re-appearing again), we have also studied the impact of the loss of signal
for a certain time on the detection. Finally, a region of 120$^{\circ}$ in
longitude centered at the middle position of the map (similar to the real
farside maps described in Sect. \ref{sect:farside_maps}) is extracted for each
time step.

\begin{figure}[!ht] 
 \centering
 \includegraphics[width=9cm]{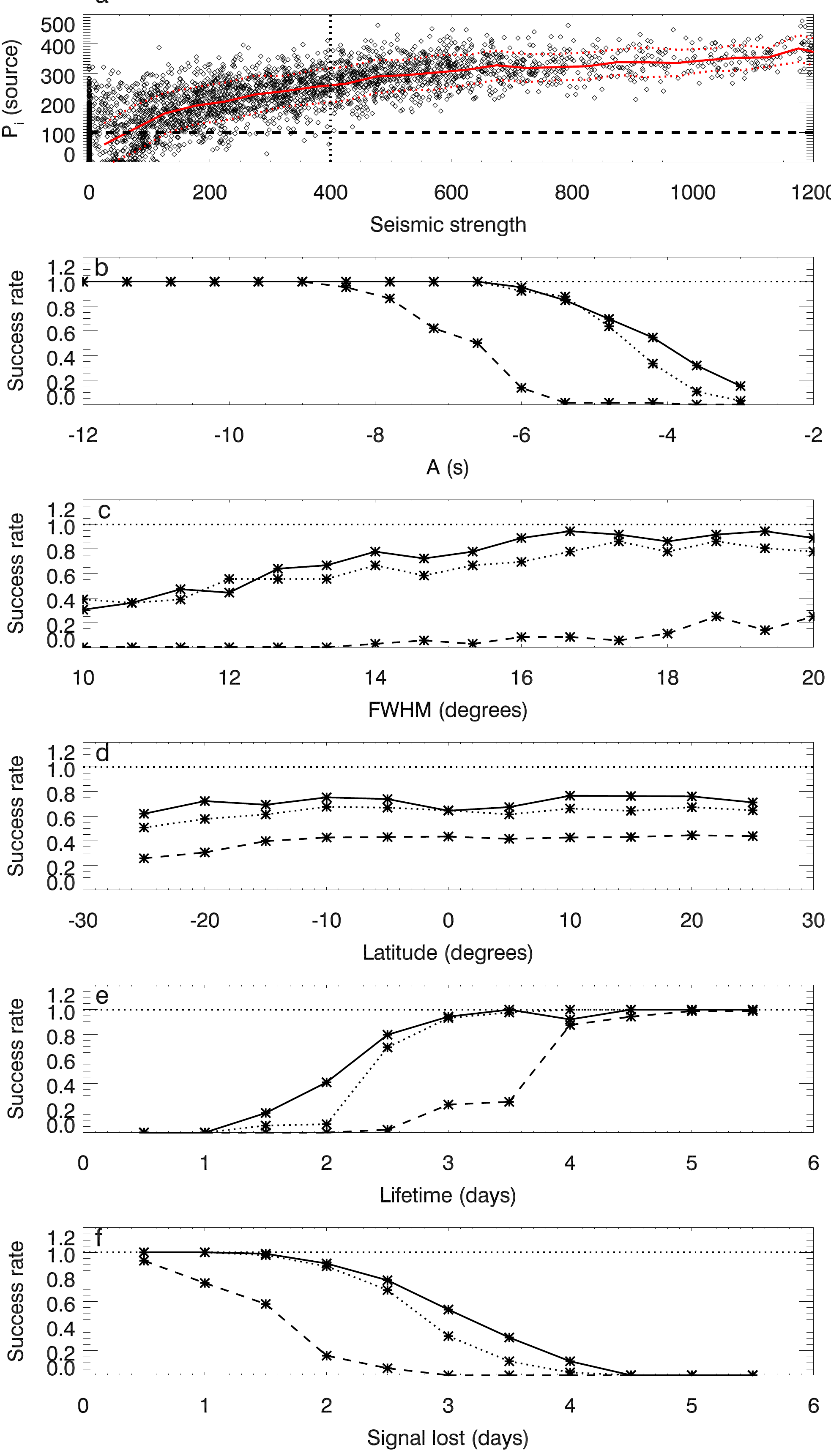}
  \caption{Analysis of 4048 artificial farside maps. Panel a shows the integrated probability of the artificial acoustic source as a function of the seismic strength. The vertical black dotted line is the threshold for the identification of a farside active region based on its seismic strength, whereas the horizontal black dashed line is the threshold for the detection of an active region using the neural network. The red solid line shows the integrated probability averaged in bins with a width of 50 in seismic strength. The red dotted lines illustrates the standard deviation of those averages. The rest of the panels shows the dependence of the success rate with the amplitude of the acoustic sources (panel b), their size (panel c), their latitude (panel d), their lifetime (panel e), and the number of seismic maps where the acoustic signal is lost (panel f). In panels b-f, the solid line with asterisks illustrates the success rate of the neural network, the dashed line with asterisks the success rate of the traditional method with a standard threshold of $S=400$, and the dotted line with asterisk is the success rate of the traditional method with a threshold of $S=65$.}      
  \label{fig:tests}
\end{figure}

Figure \ref{fig:11maps} shows an example of the 11 seismic maps constructed for
a single artificial case. In this example, $A=-9$ s, $\mathrm{FWHM}=15^{\circ}$, and the
latitude is $15^{\circ}$. For the
artificial cases we employed the same latitude and longitude coverage used for
the training. In most of the maps the acoustic source is visible as a
region with negative phase shift. As the time increases from panel a to panel k,
it is displaced in longitude due to the solar rotation, approaching to the east
limb. In some of the time steps the acoustic source is completely masked by the
noise (\eg, panel h and i).       

The top panel of Fig. \ref{fig:prediction_ejemplo} shows the 5 day average of
the artificial case discussed in the previous paragraph. It is obtained after
averaging the data from panels b to j in Fig. \ref{fig:11maps}, but keeping them
in Carrington coordinate system. In this system, the acoustic source is located
at the same longitude for all time steps (in this example, at a Carrington
longitude of $-8^{\circ}$). The signature of the acoustic source stands out
above the reduced noise obtained after the 5 day average. The temporal duration
of this average resembles the temporal span currently employed for the
measurement of farside seismic maps where the detection of active regions is
reported. The retrieved noise of the averaged artificial seismic map is
comparable to that of the actual observations. The detection of farside active
regions is claimed when a region is found with a seismic signature strength
above a certain threshold. The strength $S$ is computed as the integrated phase
shift over an area where the phase shift exceeds 0.085 rad ($\approx 4$ s). With
the area measured in millionths of a hemisphere ($\mu$Hem), a farside active
region is reported when $S>400$ $\mu$Hem rad \citep[see][]{Liewer+etal2017}. The
strength of the artificial acoustic source is indicated in the lower-left corner
of the top panel of Fig. \ref{fig:prediction_ejemplo}, whereas the red contour
delimits the region where the phase shift exceeds 0.085 rad. An active region
with a seismic signature similar to that of the case represented in this
artificial map would be detected by the current approach. 

The bottom panel of Fig. \ref{fig:prediction_ejemplo} illustrates the
probability map computed by the neural network after introducing as input the
maps shown in Fig. \ref{fig:11maps}. The blue circle in both panels indicates a
region of three times the FWHM of the source around its central location. In
that region, a large patch with high probability of the presence of an active
region is found. We have defined an integrated probability $P_{\rm i}$, computed
as the integral of the probability $P$ in a continuous feature. The
identification of the features is performed with the IDL routine $rankdown.pro$,
part of the feature tracking software YAFTA\footnote{Publicly available at
\url{http://solarmuri.ssl.berkeley.edu/~welsch/public/software/YAFTA/}}. Even
though this routine is optimized for application to magnetograms, it does a good
job grouping and labeling pixels that belong to the same feature. The $P_{\rm i}$ of the artificial seismic source is shown in the bottom-left corner of the bottom panel of Fig. \ref{fig:prediction_ejemplo}.

The neural network returns a probability map with values in the range $[0,1]$. The identification of an active region is then performed by examining those probability maps, instead of directly evaluating the travel times of the farside seismic maps. The concept of ``integrated probability'' is equivalent to the ``seismic strength'' defined by the traditional method. Rather than simply look for continuous regions with strong negative travel times, an approach which is hindered by the usual strong noise of the seismic data, the neural network provides a cleaner picture of the locations where the presence of an active region is most probable. However, the probability maps usually exhibit some significant values at regions with negative travel time as a result of noise. See, for example, the small spot out of the blue circle in the bottom panel of Fig. \ref{fig:prediction_ejemplo}. The $P_{\rm i}$ of this region is 12. 

The value of $P_{\rm i}$ is given by the probability found in a continuous region and the area of that region. We have analyzed a large set of 4048 artificial farside maps, where $P_{\rm i}$ ranges between 0 and $\approx 500$ (see top panel of Fig. \ref{fig:tests}). The artificial cases analyzed just evaluate the parameter space, and include seismic signals whose size and strength are hardly found in the actual Sun. A strong detected farside active regions exhibit a $P_{\rm i}$ up to $350$.

It is necessary to define an unequivocal
criteria to decide whether a region with increased probability is claimed as an active region or not. We have chosen to define a threshold in the integrated probability as the minimum value for the detection of seismic sources, in the same way that the traditional method establishes a threshold in the seismic strength. The selection of the threshold is based on the evaluation of the artificial set of farside maps, where we know the exact location of the seismic sources. A value of $P_{\rm i}=100$ proves to be a good compromise between the success in the detection of the seismic sources and avoiding the claim of false positives. A false positive is identified when a
feature with $P_{\rm i}>100$ is found out of a region of three times the FWHM of the Gaussian perturbation (i.e., out of the blue circle in the example from Fig. \ref{fig:prediction_ejemplo}). With this criteria, 31 false positives are found in the 4048 artificial cases explored. We note that when applying the network to real data, false positives can be easily dealt with by discarding those cases where the detection does no appear consistently in successive dates at the same location.

We have performed statistics on the performance of the neural network using the set of artificial farside maps. Figure \ref{fig:tests} illustrates the
results of the analysis of 4048 artificial cases which differ in the position of the Gaussian perturbations, amplitude, size, lifetime, and number of days with the seismic signal lost. The top panel compares the integrated probability $P_{\rm i}$ of the sources as given by the network and their seismic strength $S$. There is a strong positive correlation, as shown by the $P_{\rm i}$ averaged in bins of 50 in seismic strength (red solid line). The standard deviations of those averaged (red dotted lines) do not show strong variations across the values of $S$, being slightly higher for lower $S$. The horizontal dashed line marks a value of $P_{\rm i}=100$, that is, the selected threshold for the detection of seismic sources. The vertical dotted line is the threshold currently applied for the detection of
farside active regions ($S=400$). Those lines divide the domain in four regions.
The top-right region correspond to the acoustic sources that are detected by both
approaches (32\% of the cases). The top-left part are the cases that are only
detected by the neural network (43\%), and the bottom-right region are the sources
detected only by the traditional approach (0\%). Finally, the bottom-left region includes
weak acoustic sources that neither the neural network nor the traditional approach can identify (25\%). The acoustic
sources analyzed in this figure are just sampling a certain range in phase shift
amplitude ($-3$ to $-12$ s), size ($\mathrm{FWHM}=10-20^{\circ}$), and lifetime or number of days when the acoustic signal is lost (0.5 to more than 5.5 days). We made no
effort to reproduce the distribution of seismic signals from active regions in
actual observations.   

Panels b-f from Fig. \ref{fig:tests} compare the performance of the neural network (solid line with asterisks) and the traditional method (dashed line with asterisks) as a function of several parameters. They illustrate the success rate of the methods. For the neural network the success rate is defined as the ratio of the cases identified with $P_{\rm i}>100$ at the known location of the
source to the total number of cases. The same definition
is applied for the traditional approach, but using $S>400$ as the criteria for a
positive detection.

Figure \ref{fig:tests}b shows the success rate as a function of the amplitude of the sources, as given by a set of 1056 artificial cases. The FWHM of all the artificial cases included in the analysis is 15$^{\circ}$, and the acoustic sources are present in the 11 consecutive maps that compose each case. The sources differ in their location
(longitude and latitude) and $A$. The traditional approach can detect almost all
the acoustic sources with an amplitude above 9 s. However, its success rate is
reduced to a 50\% around $A=-7$ s and sources with $A$ below $-5.5$ s are not
detected at all. The neural network exhibits a perfect success rate for all the sources with $A$ stronger than $-6.5$ s. For $A\approx-5$ s the success rate is 50\%, and even some cases with $A=-3$ s are detected (10\% of the cases). 

\begin{figure*}[!ht] 
 \centering
 \includegraphics[width=17cm]{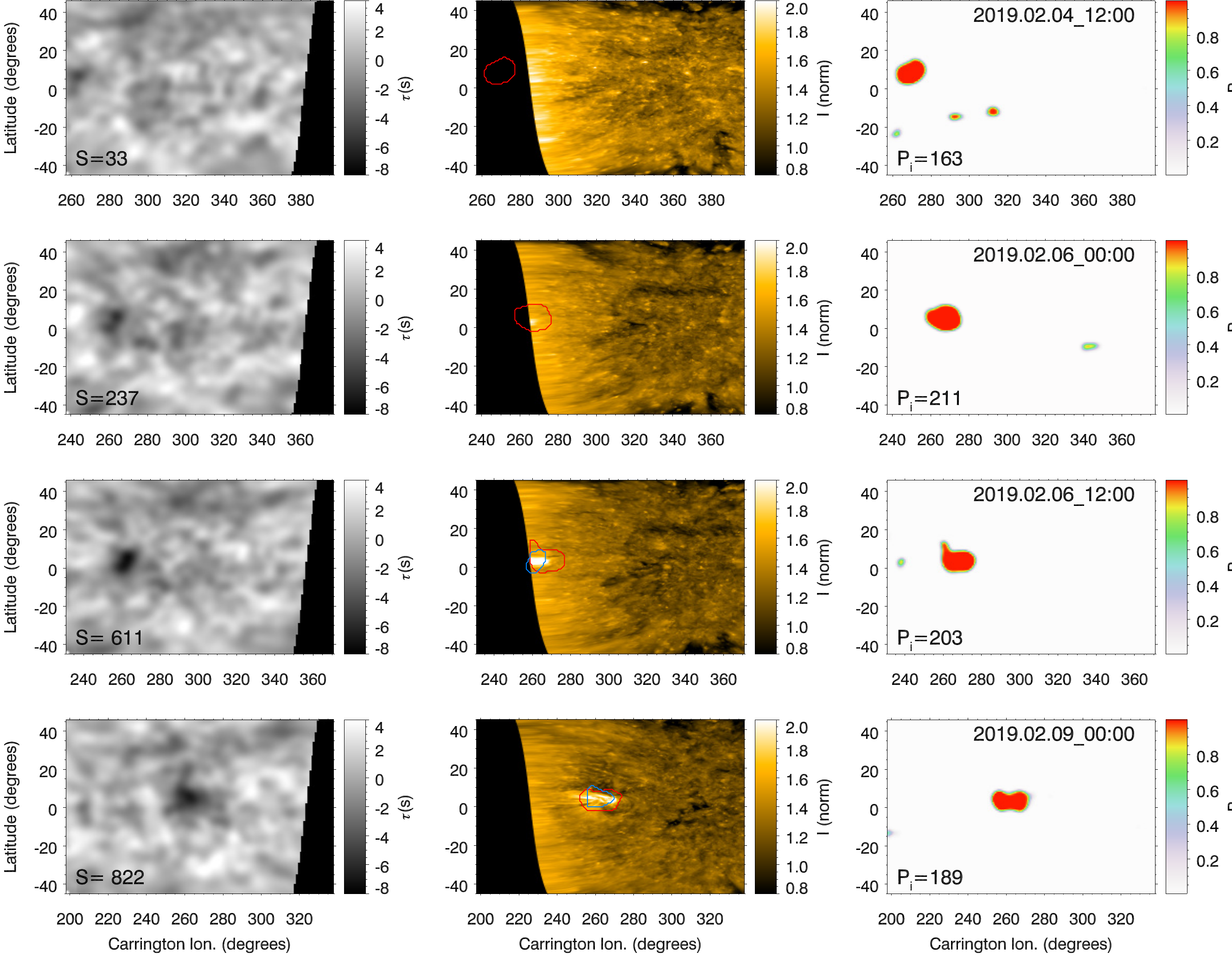}
  \caption{Detection of the farside active region NN-2019-003 (FS-2019-001). Left column: farside phase-shift maps obtained from 5 days of HMI Doppler velocity data. Bottom left of the panel shows the seismic strength of the strongest feature. Middle column: STEREO 171 \AA\ data. Color contours indicate the active regions detected by the neural network (red) and by the traditional approach (blue). Right column: Probability map, obtained as the output of the neural network. Bottom left of the panel shows the integrated probability of the strongest feature. Each row corresponds to a different time, indicated at the top part of the right panels.   }      
  \label{fig:example1}
\end{figure*}

We have also evaluated the performance of the neural network for the
identification of acoustic sources with different sizes. Another set of 1056
artificial cases has been constructed, but in this analysis all of them have the
same amplitude $A=-5$ s and they differ in the FWHM and their location in
longitude and latitude. The Gaussian signal is present in the 11 seismic maps
employed for each case. Figure \ref{fig:tests}c shows the success rate as a
function of the size of the acoustic source. The performance of the neural
network is again outstanding in comparison with the traditional approach. A FWHM
of 14$^{\circ}$ is required to start to detect some sources with $S>400$ in
the classical approach, and a
success rate of 20\% is found for $\mathrm{FWHM}=20^{\circ}$. The neural network reaches
a higher success rate for acoustic sources with half that size, and manages to
get an almost total success for sources as small as $\mathrm{FWHM}=16^{\circ}$. 

Figure \ref{fig:tests}d illustrates the efficiency of the neural network as a
function of the latitude of the sources. The rest of the parameters of the set
of cases included are the same. Their amplitude is $A=-9$ s and
$\mathrm{FWHM}=15^{\circ}$, and in some cases the signal is not present in all the individual maps. For these sources, the success rate of the network is around 80\%, although it shows
some dependence with the latitude. At the solar equator and for latitudes higher
than $\pm20^{\circ}$ the performance of the network is slightly poorer. This is
expected, since the training set is constructed with actual solar data, and few
active regions appear out of the activity belts. The network requires a stronger
signal to confirm the presence of active regions at those latitudes where they
barely emerge. The success rate of the neural network does not depend on the
longitude of the active region (not shown in Fig. \ref{fig:tests}), as far as
the seismic source is present in all the individual maps employed for the
inference.   

In the last two panels, we explore the performance of the neural network when
the signal is not present for all the 11 maps that compose each case (6 days of
data with a cadence of 12 hours). In both cases, the acoustic sources have the
same properties ($\mathrm{FWHM}=15^{\circ}$ and $A=-9$ s). In Fig. \ref{fig:tests}e we
have checked the efficiency of our model detecting active regions whose lifetime
is shorter than the 6 days of data used as input for the network. The sources
are introduced continuously in some of the 11 maps that compose each case,
ranging from a lifetime of half a day (one map) to more than 5.5 days (all
maps). It shows that the neural network can detect almost all the active regions
whose lifetime is at least 3 days, and it can even detect 15\% of the active
regions than only last one day and a half. The traditional method employs the
Doppler data from 5 days, and the signal from these short-lived active regions
is smeared out in those seismic maps. One of the main advantages of the network
is the use of series of seismic maps computed over 24 hours each, allowing us to
keep the identity of signals with short lifetime while enhancing their signature
through the incorporation of multiple days.  In Fig. \ref{fig:tests}f, we
explore the performance of the network when the seismic signal is lost in a
certain amount of nonsuccessive maps (as opposite to panel e, where the signal
disappears for successive maps). The success rate fall below 50\% when there is
no acoustic signal for more than 3 days. 

In the previous paragraphs we have discussed the comparison between the performance of the neural network using the selected threshold of $P_{\rm i}=100$ and the traditional approach using a threshold of $S=400$, which is the value currently employed in standard analyses of farside seismic maps. This evaluation is conditioned to the selection of the thresholds and, obviously, a lower threshold will offer better performance (with increased risk of false positives). The dotted lines with asterisks in Fig. \ref{fig:tests}b-f illustrate the success rate obtained from the traditional method but using a threshold of $S=65$ (the seismic strength where the red line in Fig. \ref{fig:tests}a intersects the threshold selected for $P_{\rm i}$), instead of the standard $S=400$. Their comparison with the neural network shows that the later is still superior. The neural network exhibits a higher success rate in the cases where the seismic signal is not present during all the dates (panels e and f) and for extended sources ($\mathrm{FWHM}>12^{\circ}$) with low amplitude (panels b and c). Further analyses, based on the analysis of observational data, are required to determine the thresholds that optimize both approaches.

\begin{figure*}[!ht] 
 \centering
 \includegraphics[width=17cm]{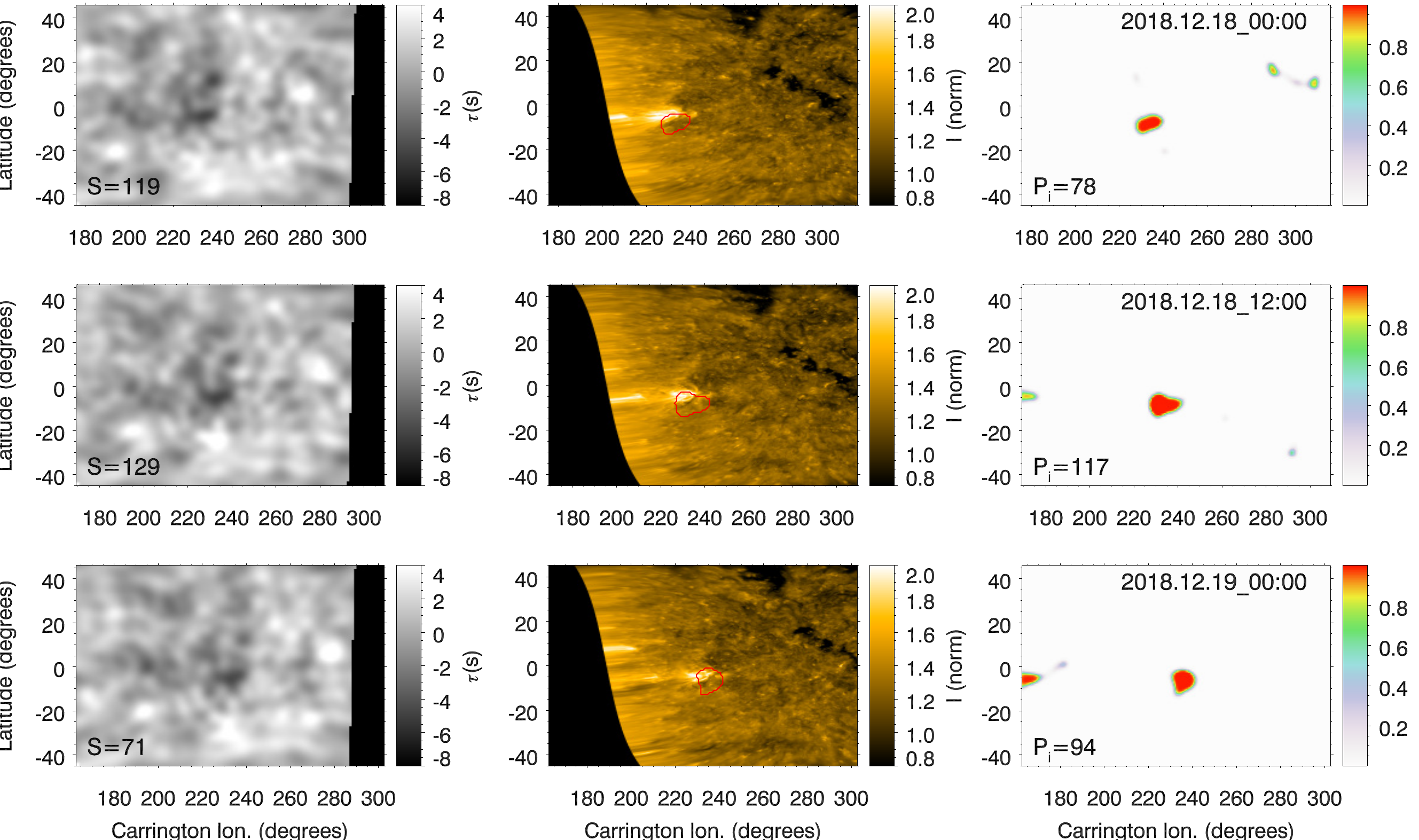}
  \caption{Detection of the farside active region NN-2018-003. Same description as Figure \ref{fig:example1}.   }      
  \label{fig:example2}
\end{figure*}

\section{Application to solar data}
\label{sect:observations}

We have applied our model to actual farside seismic maps measured between
November 2018 and May 2019, out of the period employed for the training of the
neural network. The predictions of the network have been compared with the
inferences of the traditional approach and, when available, with the EUV
emission (171 \AA\ passband) in the farside hemisphere acquired by STEREO-A
spacecraft, since magnetized regions exhibit increased brightness in the EUV.
STEREO data have previously been employed to test the reliability of farside
seismic maps for detecting strong active regions
\citep{Liewer+etal2014,Liewer+etal2017}, and a deep learning method has been
developed to retrieve solar farside magnetograms from those EUV data
\citep{Kim+etal2019}. During this time period, STEREO-A only covers
partially the hemisphere which is non-visible from the Earth. Table
\ref{table:farside_ARs} shows a list of the farside active regions detected. An
active region is claimed when a feature in the probability map exhibits $P_{\rm
i}>100$ and it appears with significant $P_{\rm i}$ at the same Carrington
longitude at least in another prediction from the neighboring dates. A total of
11 active regions have been detected in that period. The three strongest were
also identified by the traditional approach, and their counterpart in EUV
emission is found in STEREO data. From the 8 active regions exclusively detected
by the network, the signature of 5 of them is also verified by STEREO data. The
other 3 cases are detected out of the field-of-view (FOV) of STEREO, and no
signal is found when they rotate into the region observed by the spacecraft.
They possibly decayed before they were visible. The features that show $P_{\rm
i}>100$ but do not appear in neighboring predictions are considered false
positives. Five of them are found in the 353 days explored (1.4\%), similar to
the percentage of false positives expected from the analysis of artificial data.

Table \ref{table:farside_ARs} illustrates several of the properties of the
detected regions, including the given name, the date of their first detection,
their NOAA designation at the visible side, and the number of days detected. For
the later, only those detections above the thresholds (both for the neural
network and the traditional method) are considered. Note that in both approaches
the identification can be extended by tracking the same location, even if the
signature of the active region is below the threshold. Our results show that in
the case of strong active regions (those that are detected by the traditional
method), the neural network can identify them significantly earlier. Two of the
cases were detected two days in advance, while the third case (NN-2019-004) was
out of the region covered by the network and it was identified only half day
earlier. This is illustrated in Figure \ref{fig:example1}, which shows the
temporal evolution of the detection of the active region NN-2019-003. In
addition, in all those cases the signal remains longer above the identification
threshold for the neural network. 

Our model can also detect a significant amount of active regions that are missed
by the traditional approach. Figure \ref{fig:example2} shows one of those cases
(NN-2018-003). At the location of the active region, the seismic map exhibits a
slightly negative phase shift. However, its strength is not enough to claim a
detection, since non-magnetized regions show a similar phase shift (e.g.,
latitude=10$^{\circ}$ and Carrington longitude=237$^{\circ}$ at the top left
panel). We note that the use of a lower threshold $S=65$ in the traditional method, as that discussed in the previous section, would lead to a false positive.  In contrast, our model unambiguously detects the active region at its
right location, as confirmed by EUV data from STEREO-A.

\begin{table*}
\centering
\caption{Summary of the farside active regions detected in the period Nov 2018-May 2019}
\label{table:farside_ARs}
\medskip
\begin{tabular}{cccccc}
\hline\noalign{\smallskip}
Name 		    	&	Date first detection	& Number of days with      	&  AR number
& STEREO 		\\
		    	&				& signal above threshold      	&  	    	& 		\\
(NN top,        	& 	(NN top,            	& (NN top, 	            &  (previous
rot. top, 	& 	    		\\
 FS bottom)         & 	FS bottom)	           	& FS bottom)	       	&  later rot.
 bottom)	&       		\\

\hline
NN-2018-001 		& 	2018/11/23.0		&  	1.0		& 	-		& Out of	&	\\  
& 	-		&  	-	&	& FOV	&	\\

\hline
NN-2018-002 		& 	2018/12/08.0		&  	2.0		& 	-		& Confirmed	&	\\ 
& 	-		&  	-		& 	&	\\			
			
\hline
NN-2018-003 		& 	2018/12/18.5		&  	0.5		& 	-		& Confirmed	&	\\  
& 	-		&  	-		& 	&	\\			
			
\hline
NN-2019-001 		& 	2019/01/16.5		&  	0.5		& 	-		& Confirmed	&	\\ 
& 	-		&  	-		& 	&	\\			
			
\hline
NN-2019-002 		& 	2019/01/26.0		&  	1.5		& 	-		& Out of	&	\\ 
& 	-		&  	-	&	& FOV	&	\\

\hline
NN-2019-003 		& 	2019/02/04.5		&  	5.5		& 	12733		& Confirmed	&\\ 
FS-2019-001		& 	2019/02/06.5		& 	4.5		&  	-		&	&	\\

\hline
NN-2019-004 		& 	2019/03/28.5		&  	10.5		& 	12736		& Confirmed	&\\ 
FS-2019-001	    	& 	2019/03/29.0		& 	7.5		&  	12738		&	&	\\

\hline
NN-2019-005 		& 	2019/04/21.5		&  	0.5		& 	-		& Confirmed	&\\ 
& 	-		&  	-		&	&	\\
\hline
NN-2019-006 		& 	2019/04/28.5		&  	0.5		& 	12738		& Confirmed	&\\  
& 	-		&  	12740		& 	&	\\
\hline
NN-2019-007 		& 	2019/04/29.0		&  	7		& 	12739		& Confirmed 	&\\  
FS-2019-002		& 	2019/05/01.0		& 	2.5		&  	12741		&	&	\\

\hline
NN-2019-008 		& 	2019/05/09.5		&  	0.5		& 			& Out of&	\\ 	
& 	-		&  	-	&	& FOV	&	\\

\hline
\end{tabular}

\begin{tablenotes}
\small
 \item {The first column shows the label assigned to the farside active region, the second column is the date of the first detection, the third column is the number of days when the signal is detected with a value above the chosen thresholds, the fourth column is the NOAA number assigned to the active region in the previous or the following rotation, and the last column indicates the presence or absence of the farside active region in the field-of-view of STEREO-A. Each detected farside active region is indicated in two rows, the top one with the data from the detection by the neural network (noted as NN-yyyy-id) and the bottom one with the data from the detection by the traditional method (when available, noted as FS-yyyy-id).}
\end{tablenotes}

\end{table*}

\section{Discussion and conclusions}
\label{sect:conclusions}

The measurement of the magnetic activity in the farside hemisphere has multiple
applications for solar physics and, specially, for space weather forecasting.
During the last years, NASA STEREO spacecrafts have been monitoring the farside
of the Sun, providing, among other data, EUV images of that hemisphere.
Recently, \citet{Kim+etal2019} have developed a deep learning method to retrieve
solar farside magnetograms from those EUV data. However, STEREO spacecrafts are
currently returning to the Earth-side of their orbit, and there are no
guarantees that they will be operative ten years from now, when they will be
back at the farside, as contact with STEREO-B is already lost. Thus, there are
no prospects for using STEREO data for obtaining farside images in the future.
In the next years, only the ESA mission Solar Orbiter (to be launched in 2020) will
provide direct imaging of the farside, but only during some periods of its
orbit. Due to the importance of the farside magnetism for solar studies and
space weather predictions, one would expect that in the future there will be
telescopes permanently observing the whole Sun. While this future arrives, the
only method capable of constantly monitoring the solar farside is
helioseismology.     

In this work, we have developed a new methodology to detect farside active
regions from helioseismic data. We have trained a neural network using pairs of
farside maps and HMI magnetograms obtained when the helioseismically probed
region has rotated into the visible hemisphere\footnote{The neural network
can be downloaded from the repository \texttt{https://github.com/aasensio/farside}}. 
Our results show that this
method reduces the threshold in the strength of the seismic signal required to
detect it. We are able to identify smaller active regions, which produce lower
shifts in the phase, and also to detect active regions with shorter
lifetime or whose signature is lost in some of the farside seismic maps. This
allows a significant increment in the number of identified farside active
regions. 

Previous works have shown the benefits of including the farside magnetism as
input in the forecast of several data of interest for space weather, such as the
solar spectral irradiance and the solar wind \citep{Fontenla+etal2009,
Arge+etal2013}. The identification of large active regions days before they
rotate into the visible solar hemisphere is also relevant, since they can
generate sudden enhancements in the EUV irradiance at the Earth just after
appearing at the eastern limb and they also suppose a threat for solar flares. 

The analysis of the seismic signatures of farside active regions has some
limitations regarding the inference of the farside magnetism. The phase shifts
produced by active regions and measured by helioseismology is mainly produced by
the Wilson depression of the sunspots, so they are independent of the magnetic
polarity. One could try to infer the magnetic flux, but not the sign of the
polarity. It can only be guessed following the Hale's law, which correctly
predicts polarity approximately 90\% of the times \citep{Li+Ulrich2012}.
However, inferring the magnetic flux is also a challenge.
\citet{GonzalezHernandez+etal2007} tried to calibrate the magnetic flux of
farside active regions as a function of their seismic signatures. They found a
positive correlation (the higher the seismic signal, the higher the magnetic
flux), but this correlation is quite poor, which inhibits a proper determination
of the farside magnetic flux based on the measured phase shift. In this paper,
we have avoided this limitation by focusing on the determination of the
probability of the presence of an active region at a certain location of the
farside hemisphere, without associating it to the magnetic flux. Future efforts,
exploiting the capabilities of neural networks, should lead to a proper
quantification of farside magnetic flux from the analysis of seismic data.      

An obvious improvement on our approach is to completely overcome
the use of seismic maps and work directly with Doppler maps. Such
an approach could potentially lead to the development of a data-driven 
farside helioseismological method, which could better exploit the information 
encoded on the Doppler maps. We anticipate that this would
require an architecture that is able to deal with very long time series. Note
that the seismic maps used in this work have been obtained with Doppler information
with a cadence of 45 s, so that one ends up with 1920 Doppler 
measurements per day. A possibility worth exploring is the use of recurrent
neural networks with attention mechanisms, like those used in 
neural machine language translation \citep{2014arXiv1409.0473B}.

\begin{acknowledgements} 
Financial support from the State Research Agency (AEI) of the Spanish Ministry
of Science, Innovation and Universities (MCIU) and the European Regional
Development Fund (FEDER) under grant with reference PGC2018-097611-A-I00 is
gratefully acknowledged. This research has made use of NASA's Astrophysics Data System
Bibliographic Services.
We acknowledge the community effort devoted to the development of the following
open-source packages that were used in this work:
\texttt{numpy} (\texttt{numpy.org}), \texttt{matplotlib} (\texttt{matplotlib.org}),
\texttt{astropy} (\texttt{astropy.org}), \texttt{h5py} (\texttt{h5py.org}),
\texttt{scipy} (\texttt{scipy.org}),
and \texttt{PyTorch} (\texttt{pytorch.org}
\end{acknowledgements}

 \bibliographystyle{aa} 
 \bibliography{biblio.bib}

\end{document}